\documentclass[a4paper,twoside]{article}

\usepackage{epsfig}
\usepackage{subcaption}
\usepackage{calc}
\usepackage{amssymb}
\usepackage{amstext}
\usepackage{amsmath}
\usepackage{amsthm}
\usepackage{multicol}
\usepackage{multirow}
\usepackage{kbordermatrix}
\usepackage{pslatex}
\usepackage{apalike}
\usepackage{caption}
\usepackage{url}
\usepackage[hyphenbreaks]{breakurl}
\usepackage{SCITEPRESS}     

\usepackage{breakurl}

\begin{document}

\title{Enterprise Architecture Model Transformation Engine}

\author{
	\authorname{Erik Heiland\orcidAuthor{0000-0002-6636-8356}, Peter Hillmann\orcidAuthor{0000-0003-4346-4510} and Andreas Karcher
}
\affiliation{
	Universität der Bundeswehr München,\\ Werner-Heisenberg-Weg 39, 85577 Neubiberg, Germany
}
\email{
	\{erik.heiland, peter.hillmann, andreas.karcher\}@unibw.de}
\vspace*{-6mm}
}


\abstract{
With increasing linkage within value chains, the IT systems of different companies are also being connected with each other.
This enables the integration of services within the movement of \mbox{Industry 4.0} in order to improve the quality and performance of the processes. 
Enterprise architecture models form the basis for this with a better buisness IT-alignment.
However, the heterogeneity of the modeling frameworks and description languages makes a concatenation considerably difficult, especially differences in syntax, semantic and relations.
Therefore, this paper presents a transformation engine to convert enterprise architecture models between several languages.
We developed the first generic translation approach that is free of specific meta-modeling, which is flexible adaptable to arbitrary modeling languages.
The transformation process is defined by various pattern matching techniques using a rule-based description language.
It uses set theory and first-order logic for an intuitive description as a basis.
The concept is practical evaluated using an example in the area of a large German IT-service provider.
Anyhow, the approach is applicable between a wide range of enterprise architecture frameworks.
}

\onecolumn \maketitle \normalsize \setcounter{footnote}{0} \vfill

\section{\uppercase{Introduction}}
\label{sec:intro}
One major purpose of Enterprise Architecture~(EA) Models is the support of the management in controlling an organization.
With process descriptions, it is an essential toolkit for the integration and automation of processes in a value chain.
The diversity of over 50 EA Frameworks~\cite{Matthes.2011} becomes a problem when multiple stakeholders would like to work together.
Especially, if the collaborating stakeholder uses different EA frameworks, which are not direct compatible. 
Model transformation is a core technology to overcome these challenges by converting models from one framework to another.
It enables the decoupling of concrete and detailed modeling by means of the syntax of different meta-models.
Furthermore, model transformation supports analyzing models to automatically check on conceptional flaws or improve maturity level.
All in all, model transformation leads to the integration of heterogeneous languages to a uniform, homogeneous and consistent model.
This enables a model-driven EA from the highest abstraction level to the technical realization in detail via a unified tool chain.
Each abstraction layer offers their service to the upper layer whereas services of lower layers are used through a standardized interface.
However, we are currently still far away from this ideal vision to obtain a consistent overview of an enterprise with the possibility of a technical deep dive.
The development process is thwarted by description language disruption.
Furthermore, the questions on verification and validation are also open to ensure that the produced results are comprehensible.
This requires the development of new approaches that are easy and intuitive usable for the people.

In this paper, we present a generic transformation engine for EA models.
The combination of general transformation principles, rule based approach, and pattern matching is efficient in the context of model transformation management.
The design is focused on simple applicability as well as on a verifiable process.
It is not restricted to a specific modeling language and can be used for cross EA frameworks.
For this purpose, we demonstrate that the usage of our engine is intuitive, flexible, and easy to use.
Then the approach is summarized and a short outlook is given.

\section{\uppercase{Requirements}}\label{sec:require}
Consider the following scenario, a supplier of car parts is merged in an automotive manufacturer.
The existing IT systems should be integrated in the current landscape and processes for an consistent infrastructure.
The interfaces to applications and data structures are available in the form of an architecture building blocks as UML models.
Nevertheless, the modeling of processes at the automotive manufacturer and the supplier has been done in different languages and in diverse levels of detail up to now.
The automotive manufacturer used \textit{Business Process Model and Notation (BPMN)} with many details whereas the supplier used \textit{ArchiMate} only for the main aspects.

In particular, the problem lies in the several and dispersed data storages.
These are characterized by different formats (e.g. date as text or numeric), degree of detail (e.g. place as address or city) and meaning (e.g. light as spotlight or ambience).
A more difficult problem is the use of relationships to display information.
On the one hand, it is an application, and on the other hand, it must be a service based on at least one server supported by a role.
Nevertheless, the customer demands the documentation of the desired IT solution generally in the form of architecture models, similar to a formalized requirement specification.

There are several challenges in the alignment of heterogeneous frameworks: tool compatibility, language syntax, exchange formats, and semantic and expressiveness of the meta models.

In this paper, we focus on providing a solution to address semantic heterogeneity, without depending on a single tool or framework.

Considering the manifold of EA modeling possibilities, we focus on a wide applicable approach.
For a generic transformation language, the following requirements have to be considered:
\begin{itemize}
\item Easy applicability by intuitive and several description possibilities
\item Separation of content description, rule definition, and tool environment to avoid dependencies
\item Complete functional scope for conversion with an independent transformation engine for a wide adaptability
\item Chaining and interlinking of transformation rules
\end{itemize}
Beside these main requirements, there are further minor demands.
A transformation language has to avoid the usage of explicit meta-models to be flexible adaptable as a generic approach, and to focus on the content to be converted.
The meta-model is inherent included in the model anyway.
The syntactical correctness of the generated target model in relation to a meta-model is independent of the transformation itself. It is checked afterwards.
Furthermore, a transformation language fulfilling the commutative and associative property allows a bijective transformation and compatibility.
\section{\uppercase{Related Work}}
\label{sec:related_work}
In the literature, there are over 60 tools for model transformation, which have been analyzed in various surveys \cite{Kahani.2018,NafisehKahani.2015,Jakumeit.2014}.
Most of them are not available any more.
Furthermore, existing tools are not usable in practice or in industrial context, because these are too complex and mainly meta-model specific for academic purpose.
Current approaches for the problem can be divided in three categories \textit{Enterprise model Transformation}, \textit{Enterprise ontologies}, and \textit{Ontology alignment}.
\subsection{Enterprise model transformation}
With regard to the technical realization of the mapping of Enterprise Architecture Frameworks we consider solutions from the field of model transformation. 
Numerous tools exist for model-to-model transformation. 
The survey of Kahani~\cite{Kahani.2018} classified 60 of them according to several criteria.
Most of them implement the OMG standard QVT~\cite{OMG.June2016} or similar, using OCL~\cite{OMG.2014} or other expression languages to define transformation rules. 
Our approach addresses domain experts in enterprise architecture modeling who do not necessarily have experience in programming, meta-modeling, or model transformation. 
Therefore, we follow the definition of Acretoaie~\cite{Acretoaie.2018} of "end-model-users", who need a solution that is as intuitive and easy to learn as possible. 
Their Visual Model Transformation Language~(VTML) is the only language known to us that pursues such an approach. However, it only allows endogenous transformations, which means that source and target models conform to the same meta-model~\cite{Westfechtel.2018}.
This conflicts with our mandatory requirement for a model transformation between heterogeneous frameworks.

\subsection{Enterprise ontologies}
Ontologies described in machine-readable form like OWL~\cite{WorldWideWebConsortium.2012b} are a suitable starting point for the semantic alignment of enterprise models.
Hinkelmann~\cite{Hinkelmann.2016} describes the advantages of describing enterprise architectures as ontologies in terms of enterprise analysis and decision making.
Beside this, there are different methods for developing ontologies~\cite{Forbes.2018b} and various projects regarding the design of enterprise ontologies~\cite{Fedotova.2018}.

We consider methods that deal with the semantic enrichment of existing enterprise models, respectively their meta-models, with regard to the improvement of the compatibility to other frameworks. 
Hadidi~\cite{AlHadidi.2019} developed a domain ontology for extended and virtual Enterprises to improve knowledge sharing within such collaborations.
The adaptation requires that the existing business models are already described as ontologies. 
Furthermore, only a small subset of possible architectural elements is provided under the concept \textit{framework}, into which the models must be classified. 
This abstraction is not sufficient for us as preparation for a semantic alignment of different frameworks.

\subsection{Ontology alignment}
The semantic coordination of heterogeneous enterprise architecture frameworks is a prerequisite for transformation at model level. 
One aspect of our research is the derivation of transformation rules by aligning the frameworks on the basis of their representation as ontologies.
This excludes a merging of both ontologies, since the result must contain the relational connections between both ontologies. 
Various general matching techniques are discussed by Ramar~\cite{Ramar.2016} and Hu~\cite{Hu.2017}. 
Furthermore, there are more than 90 tool-based approaches for (semi-)automatic matching, of which only a few are still available and applicable~\cite{Ganzha.2016}. 

Chen~\cite{Chen.2019} examined tools with a focus on visualization of the alignments, which appear to be most suitable for a practical application in the context of EA mapping.
Nevertheless, the research approaches only fulfill a subset of the corresponding relationships between two frameworks that are relevant for us. 
According to the definition of Ochieng~\cite{Ochieng.2018}, ontology matching tools serve to identify equivalent and disjunctive concepts, as well as subsumption.
For example, we still need statements about aggregation and case-based equivalence depending on existing instances.
However, the semantic integration approaches can only support the enterprise architecture transformation.

To the best of our knowledge, there is no holistic approach that meets all the identified requirements on a language independent enterprise model transformation.
\section{\uppercase{Design for Generic Enterprise Model Transformation}}
\label{sec:language_design}
From the requirements described in Section \ref{sec:require}, a solution is derived that can handle the diversity of meta-models and languages within this context. 
We focus on a generic design to bridge tool boundaries, description formats and semantic differences.

\subsection{Transformation Organization and Rule Structure}
\begin{figure}[bth]
	\centering
		\includegraphics[width=0.47\textwidth]{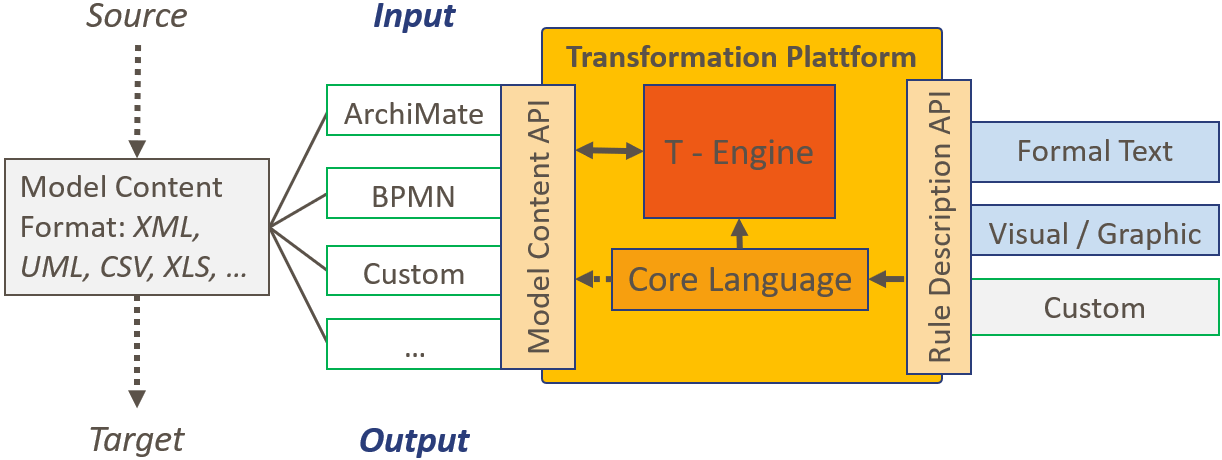}
	\caption{Transformation Engine Overview}
	\label{fig:overview}
	\vspace{-1mm}
\end{figure}

Based on the domain of EA Transformation, we have to deal with a multitude of heterogeneous meta-models.
Content is typically described in terms of elements and relations, whereby other formats (e.g. textual, tables, etc.) are also possible.
The diversity of description languages and degrees of formalization should not influence the functionality of a transformation language, which is why a decoupling is proposed at this point.
In order to different description forms for transformation rules, a variability is also provided here.
Figure \ref{fig:overview} provides an overview of our concept of a holistic approach for a generic transformation language.

\begin{figure}[bth]
	\vspace{-1mm}
	\centering
	\includegraphics[width=0.40\textwidth]{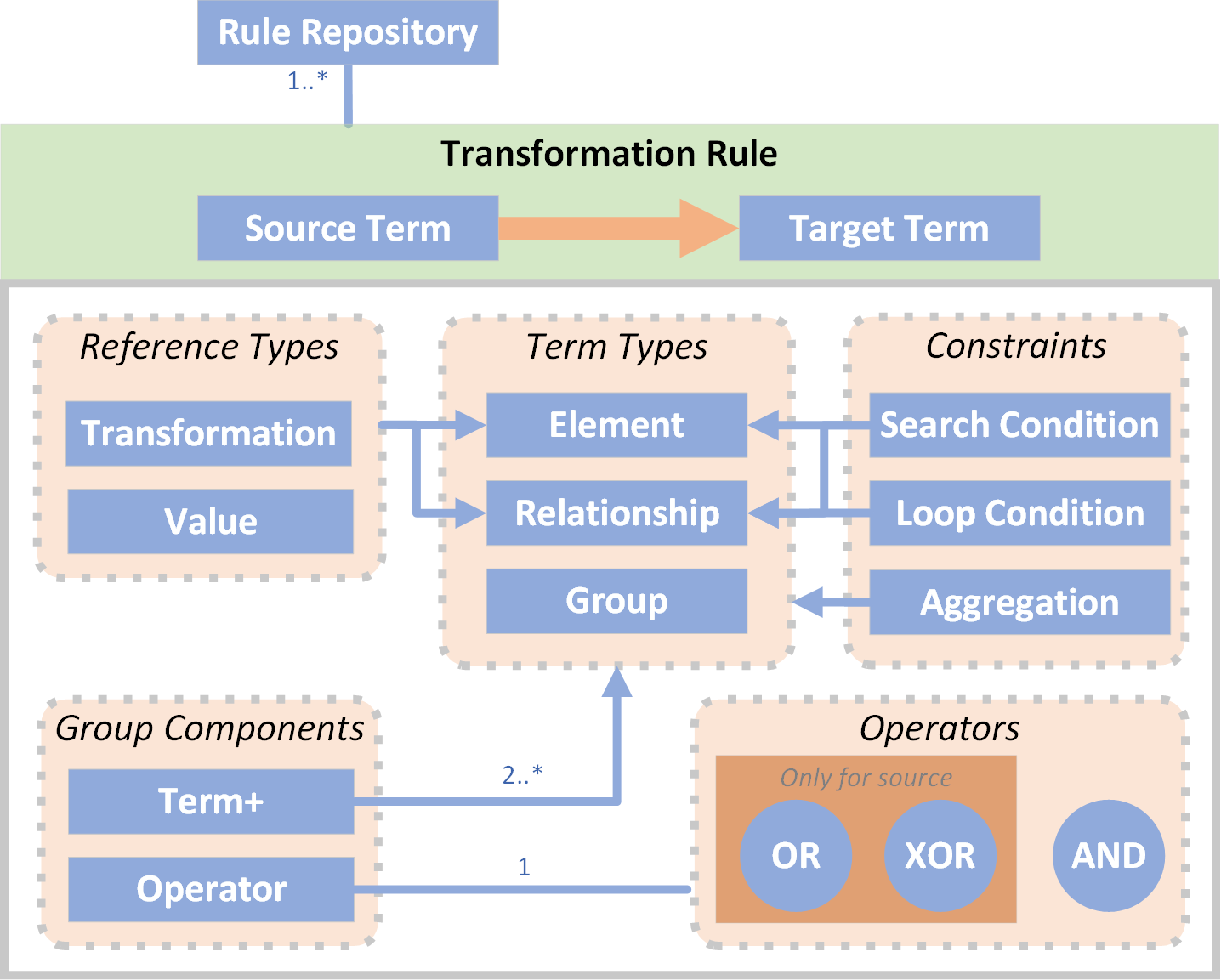}
	\vspace{1mm}
	\caption{EMT Description Components}
	\label{fig:transformation_description_components}
	\vspace{-1mm}
\end{figure}

There are many possibilities to describe an Enterprise Model Transformation (EMT).
We decided upon a rule-based approach with loose coupling. 
It allows the decomposition of complex model transformations into understandable parts. 
In contrast to coherent descriptions, these rules can be more easily reused for other transformations or replaced by others.
Based on rules, a graphical description can be invented on top of it, which provides better readability for humans.
An overview of the most important concepts of the language is presented in Figure \ref{fig:transformation_description_components}.
A \textit{Rule Set} contains 1..n \textit{Transformation Rules} and forms the basis for an overall transformation. 
Each rule is an executable unit, which can be independent or linked to other rules within a common rule set.
A rule consists of a \textit{Source Term}, which represents the input condition, and a \textit{Target Term}, which contains the build rules for the target model:

\begin{small}
\begin{verbatim}
RuleSet ::= TransformationRule*
TransformationRule ::= 'rule('Name':   
              'SourceTerm'->'TargetTerm')'
SourceTerm ::= SourceElement | SourceGroup |
               SourceRelationship
TargetTerm ::= ...
\end{verbatim}
\end{small}

\textit{Terms} are abstract constructs and can be instantiated by \textit{elements}, \textit{relations} and \textit{groups}.
Elements and relations are referred in the following as \textit{transformation items}, if statements apply to both types.
\textit{Groups} are used to encapsulate logically related statements and can be nested into each other at will.
For the logical combination of input conditions the operators \textit{AND}, \textit{OR}, and \textit{XOR} can be used.

Transformation items are used for the actual query of the model contents to be transformed.
They have a reference to a \textit{metatype} that is used to identify potential content according to its type.
If the content is not to be filtered by the metatype, a wildcard is set here using the * - operator.
\textit{Constraints} can be used to refine the statement, for example to restrict it by namespaces or existing attributes.
They also serve to aggregate statements about the entire model, which can influence the number of execution processes of the rule.

The production rules are defined within the target term.
A distinction is made between two types: creation of new elements or relations and enrichment of existing objects from other rules.
\textit{Target Groups} always represent an AND link of all statements.
\textit{Transformation references} are used to link rules with each other. 
\textit{Value references }are used to specify the target items, for example by assigning names, attributes and tags.

\subsection{Content Query Techniques}
\label{sec:patternMatching}
In the following, the primarily used query techniques are introduced in order to provide a detailed understanding of how the transformation engine works.
\subsubsection{Content Filtering and Aggregation:}
\label{sec:contentFiltering}
Each condition specified within a source term is used to query the existence of a certain constellation within the overall model.
The transformation items, which query a subset of the model by their type \textit{A}, form the basis for this.
The processing of these contents \textit{$ a_{1}..a_{n} $} takes place iteratively by default as shown in equation \eqref{matrix1}. As each rule is self-contained, deterministic results are obtained. Thus race conditions are excluded. The iterative rule matching process run until no rule matches anymore. To ensure that the transformation is executed correctly, an exact and specific description must be ensured when creating the rules, whereby the rules must not overlap.
\begin{equation}
\vspace*{-1mm}
\kbordermatrix{ & A \cr
              & a_{1} \cr
              & \vdots \cr
							& a_{n} \cr} 
							\begin{array}[pos]{ccccc}
								\rightarrow & Exec & \{a_{1}\} &\\
								 & \vdots &  &\\
								\rightarrow & Exec & \{a_{n}\} & \\
							\end{array}\\
\label{matrix1}
\vspace*{1mm}
\end{equation}

This means that whenever an item corresponding to the type exists, the rule is executed and the respective contents are forwarded to the target model generation.
Content can be further specified by additional criteria through \textit{Search Conditions}, e.g., filtering by name or some property.

If the transformation is not to be executed for every existing element or for every relationship between two elements, the contents must first be aggregated. 
\textit{Loop constraints} are introduced for this purpose.
This type of aggregation forms the iteration into an existence check condition.
This means for elements, instead of triggering an execution per element, a query is made for the existence of a number of existing elements according to the condition.
For example, it is possible to check whether at least one element of the specified type exists.
The effect on the execution of the rules are explained in Equation \eqref{eq:m_2}.
\begin{equation}
\vspace*{-1mm}
\hspace*{-3mm}
\kbordermatrix{&A \cr
              &a_{1} \cr
              &\vdots \cr
							&a_{n} \cr} \rightarrow
							\kbordermatrix{&A\{c(\geq 1)\}\cr
							&a_{1},  a_{\ldots}, a_{n} \cr} 
							\stackrel{A.Count\; \geq \; 1}{\longrightarrow}Exec\{a_{1},  a_{\ldots}, a_{n}\}\\					
  \label{eq:m_2}
  \vspace*{1mm}
\end{equation}

As a stand-alone condition, exactly one execution takes place if the statement is true.
The operands available are \textit{at least} $(\geq)$, \textit{more than} $(>)$, \textit{exactly} $(=)$, \textit{less than} $(<)$ and \textit{at the most} $(\leq)$.
Ranges can also be queried by combining several loop constraints. 

For the aggregation of source relationships there are 4 different possibilities besides the standard iteration over all relations of the type.
The configuration is achieved by setting loop constraints on the relations source (S), target (T) and the relation (R) itself.
Table~\ref{table:loopconstraints} shows the possible combinations, where 1 means that at least one loop constraint has been assigned.
\setlength{\tabcolsep}{3pt}
\renewcommand{\arraystretch}{1.3}
	\begin{table}[bth]
		\vspace*{-1mm}
		\centering
		\small
			\begin{tabular} {c|c|c|p{6cm}}
				S & R & T & Result\\ \hline
				0 & 0 & 0 & Iteration over all relationships of the type\\ \hline
				\multirow{2}{*}{0} & \multirow{2}{*}{1} & \multirow{2}{*}{1} & Iteration over distinct source elements, aggregation of  relations and target elements  \\ \hline
				\multirow{2}{*}{1} & \multirow{2}{*}{1} & \multirow{2}{*}{0} & Iteration over distinct target elements, aggregation of  relations and source elements  \\ \hline
				\multirow{2}{*}{0} & \multirow{2}{*}{1} & \multirow{2}{*}{0} & Iteration over all source - target combinations, aggregation of corresponding relations \\ \hline
				\multirow{2}{*}{1} & \multirow{2}{*}{1} & \multirow{2}{*}{1} & Aggregation of all relations, source and target elements to one statement \\
			\end{tabular}
				\caption{Different Query Patterns for Relationships.}
				\label{table:loopconstraints}
				\vspace{-3mm}
	\end{table}

\subsubsection{Logical Operations:}
If several source terms are grouped, the provided contents are combined using the defined logical operators.
For this purpose, calculations using AND, OR, and XOR are possible.
The quantity comparisons take place in pairs and refer to the names of the transformation items. 
Contents of identically named parameters are merged and checked for compliance with the statement, while differently named parameters are appended.
Finally, the valid content sets are transmitted to the target term for execution.

\subsection{Rule Concatenation}
\label{sec:references}
\begin{figure}[h]
	\centering
		\includegraphics[width=0.47\textwidth]{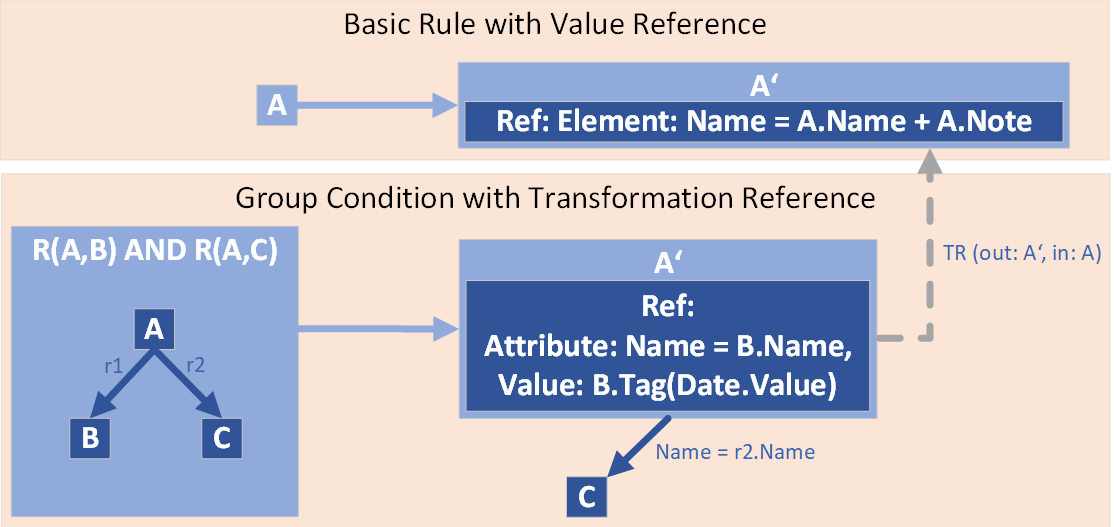}
	\caption{Mapping Example using Transformation References}
	\label{fig:mapping_concept_overview_example1}
\end{figure}

Transformation references allow the reuse of already created output items from other rules.
This allows splitting of complex model areas into partial transformations and incremental enrichment.
As shown in Figure~\ref{fig:mapping_concept_overview_example1}, the target element is created in one rule and enriched with attributes and relations in another.

The referenced rule works like a function that returns the specified target values, based on the input parameters passed to it.
If the called rule has not yet been executed, this occurs at the latest when the function is called in order to generate its value range.
Depending on the definition of the rules and the input parameters passed, a transformation reference can return zero, exactly one or more values for the application of further transformation steps.

\begin{figure}[]
	\centering
		\includegraphics[width=0.47\textwidth]{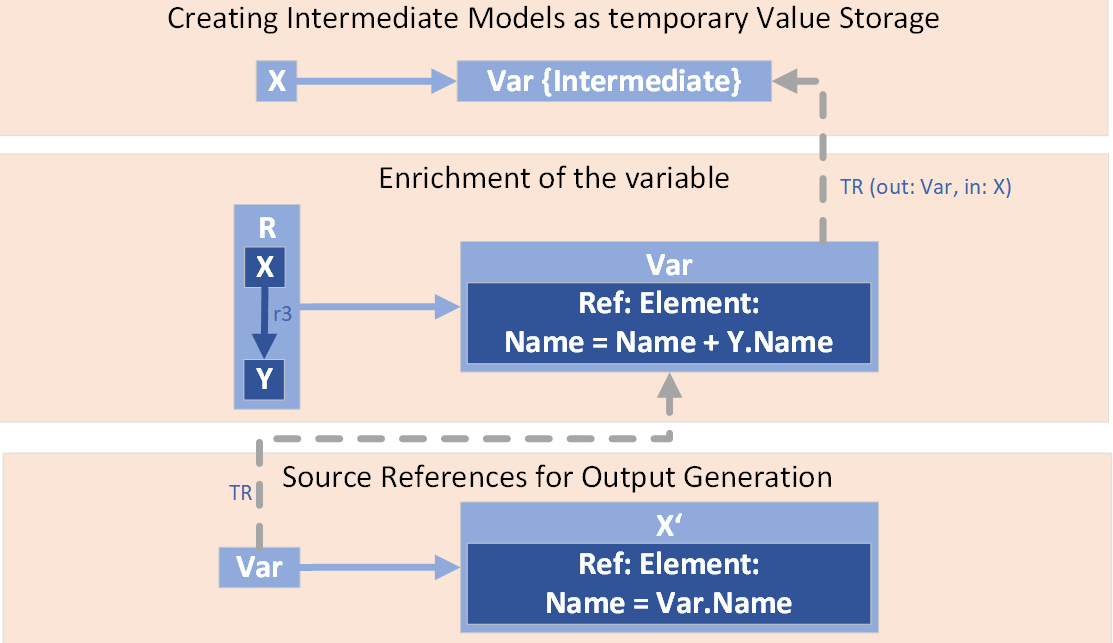}
	\caption{Transformation Reference with Intermediate Elements}
	\label{fig:mapping_concept_overview_example2}
\end{figure}

They can also be used on the source side of a rule.
This mechanism can be used, for example, to concatenate transformation rules across multiple frameworks.
It is also possible to mark certain elements and relations as intermediate.
Thus, they act as variables, which become accessible for further rules by references.
These elements are removed at the end of the transformation.
Figure~\ref{fig:mapping_concept_overview_example2} shows an example, how a variable is enriched step by step with a name and how this variable is finally used to create an element.
Such structures can be useful, for example, to convert a tree structure into a flat one, such as a pivot table.

\subsection{Meta Model Assignment}
The advantage of the presented EMT language is that it allows the definition of mapping rules without explicitly described meta models. 
Thus, no profile or schema is required, types can be simply declared and assigned to rules.
A connection is made via the name, which must also be assigned to each content element in the form of the type.
The assignment of several meta types for content objects is possible. 
This can be useful for more aggregated considerations. Table \ref{table:metatypes} shows an example of how elements can be assigned to the higher-level types in addition to the direct metatype.
	\setlength{\tabcolsep}{1pt}
	\begin{table}
	\centering
	\footnotesize
	\begin{tabular} {l | l | l }
		Content & Metatype 1 & Metatype 2  \\ \hline
		Customer & BusinessActor & Business (A) \\
		Financial App. & ApplicationComponent & Application (A) \\
		Payment Proc. & ApplicationService & Application (B)  \\ 
		Rep. DB Updates & TechnologyService & Technology (B)\\ 
		Quoted Price & DataObject & Application (P) \\ 
		r1 & AggregationRelationship & Relation (E)\\ 
		r2 & CompositionRelationship & Relation (E)\\
	\end{tabular}
	\caption{Assignment of multiple metatypes based on elements of ArchiMate~\cite{.2016}}
	\label{table:metatypes}
\end{table}
	
These overlapping results offer several possibilities for model queries.
For example, the following execution conditions can be formulated: 
\begin{itemize}
	\item for all elements from type\\ 'BusinessActor':  element(A:BusinessActor)
	\item for all active structure elements within the model:\\ element(A:Active)
	\item for all behavior elements from application Layer: \\(element(A:Behavior)$\wedge$(element(A:Application)))
\end{itemize}

The concept is also suitable for other languages where type inheritance is possible. 
This is the case, for example, with UML profiles where specialized stereotypes are derived from meta classes.
In addition to aggregation, the multiple assignment can also be used to define synonyms to compensate potential format differences, like "archimate:BusinessActor" and "Business Actor".	

\subsection{Adaption of Content- and Rule Description Interpreters}
The generic structure allows the separation of the actual transformation from the description format of the source and target model, as well as the rule description. 
To formalize the contents according to the syntax of the transformation language, interfaces are defined that can be implemented by developers with little effort.
Thus, the transformation language can be used for each individual EA and domain-specific language (DSL).

\subsubsection{Transformation Content:} 
The transformation language processes content that must be prepared according to a defined interface.
A distinction is made between entities and relations.
Both define attributes for name, id, and metatypes. In addition, relations contain references to their source and target entities.
	
The \textit{GetValue(expression)} function is used by the transformation engine to address additional object-specific attributes for creating and enriching target objects using value references.
The definition of the syntax used for this can be determined by the user.
The second function \textit{SetValue(referenceResult)} is used to generate the target models from the generic output format of the transformation.
A variety of evaluations are also possible here, depending on the specified value references. 

\begin{figure}[bt]
	\centering
		\includegraphics[width=0.47\textwidth]{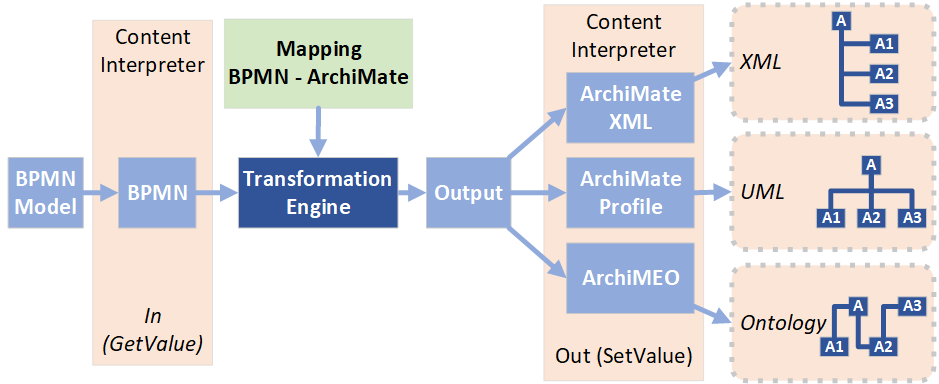}
	\caption{Using different Content Interpreters for converting the Transformation Output}
	\label{fig:different}
\end{figure}

For example, the transformation of process models from BPMN \cite{ObjectManagementGroup.2013} to ArchiMate is realized independently of the target language.
Afterwards, a different content interpreter can be used to import the content or to generate a model according to the ArchiMate Model Exchange Format \cite{TheOpenGroup.2017}, an UML profile for ArchiMate or the ArchiMEO ontology \cite{Hinkelmann.2013} (see Figure \ref{fig:different}).
	\subsubsection{Rule Description:} 
	The transformation rules can be defined using the provided grammar according to fist-order logic and made available to the transformation engine.
	The mapping of language constructs to model elements is simple due to the clear structure.
	Therefore, different visual notations can be further applied, which allow the description of the mapping rules in arbitrary modelling tools.
	Thus, an enterprise architect can model the transformation rules in its familiar environment.

\subsection{Transformation Management}
The mapping of complete EA frameworks can quickly contain over 100 chained transformation rules.
Therefore, grouping as well as monitoring and debugging is necessary to ensure rule consistency and correct execution.
Our approach in the direction to transformation management includes the following aspects:
\begin{itemize}
	\item Dependency Check: The relationship between individual rules must be comprehensible. 
	This ensures that rules do not refer to each other, generating endless loops, and content is not mapped redundantly and further processed at different points.
	\item Content Matching: It must be comprehensible how the query mechanisms work on the model and which content is transformed.
	For this purpose, it must also be possible to make partial queries to the individual source terms in complex rules as well as single rule execution.
	\item Traceability: After the transformation it must be possible to determine which rules were responsible for which target content under consideration of which content.
\end{itemize}
Especially the last aspect is relevant for an incremental transformation.
Our approach provides traceability via assignment tables.
Each rule maintains a directory of inputs and outputs as well as their execution status.
An already created object can be reused and adapted exactly as described in Subsection \ref{sec:references}.
The final transfer of the change from this generic transformation output format to the target format must be the task of the content interpreter.
The prerequisite is that the reference between target and output format, is retained during interpretation, for example via the global identifiers of the objects.

\section{\uppercase{Proof of Concept}}
\label{sec:evaluation}
For demonstration of the transformation language, we applied a graphical syntax in the area of business process modeling.
Therefore, we developed a prototype in C\# as a plug-in for Enterprise Architect from SparxSystems, which allows the definition and execution of transformation rules.
In the following example, we show an excerpt of the mapping approach between BPMN and ArchiMate to demonstrate the expressiveness and applicability of our solution, listed in Table \ref{table:mapping_archi_bpmn}.
	\setlength{\tabcolsep}{2pt}
	\renewcommand{\arraystretch}{1.1}
	\begin{table}
		\centering
		\small
			\begin{tabular} {l | l }
				 ArchiMate & BPMN   \\ \hline
				 Business Actor, Role & Participant/Pool, Lane \\
				 Application Component & Participant/Pool, Lane \\
				 Business & Collaboration\\
				 Application Collaboration & Collaboration\\
				 Business/Application Process & Process \\
				 Triggering & Sequence flow\\
				 Access & Data association\\
			\end{tabular}
				\caption{Aligning ArchiMate with BPMN \cite{Lankhorst.2017b} }
				\label{table:mapping_archi_bpmn}
	\end{table}
	
We focus on the transformation of different facts and omit the definition of value references for better clarity. 
These references are placed using annotations or attributes, as shown in previous Figures~\ref{fig:transformation_description_components},~\ref{fig:mapping_concept_overview_example1}, and~\ref{fig:mapping_concept_overview_example2}.

\begin{figure}[]
	\centering	\includegraphics[width=0.47\textwidth]{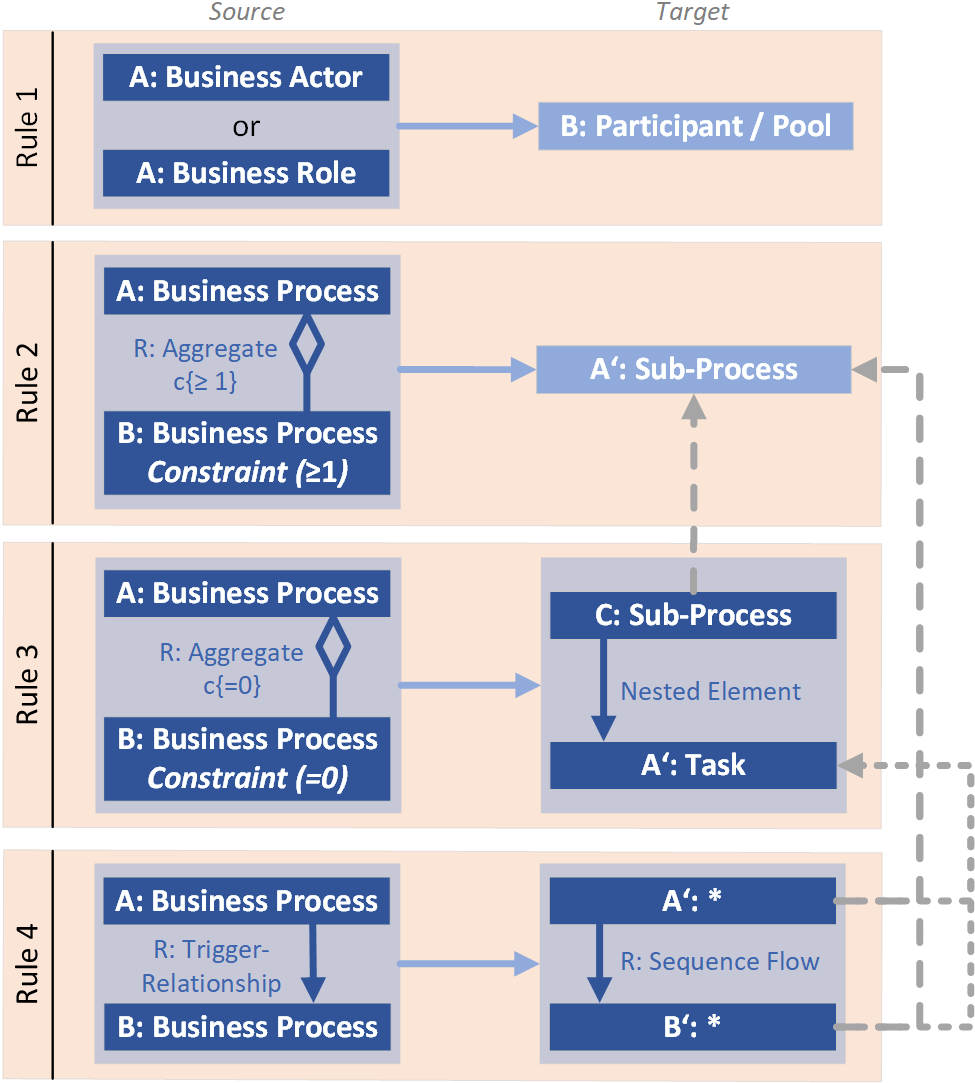}
	\caption{Demonstration Example of the application of EMT}
	\label{fig:demo}
\end{figure}

The described transformation of processes, links and actors is presented in Figure~\ref{fig:demo}.
This short scenario shows that only a few language building blocks are necessary to form the model expressions to map between EA models.
According to our scenario, the supplier processes could be integrated into the automotive manufacturer's process model. 
The integrated process landscape ensures coordinated processes and a uniform overall understanding for such an extended enterprises~\cite{AlHadidi.2019}. 
Business changes could be communicated through incremental model adjustments and responded to complications accordingly.

First, we map all Business Actor and Business Role elements to Participant/Pool elements (Rule 1). 
Therefore all elements from type actor or role are combined in a source group under the parameter A and transformed to an element B with type 'Participant / Pool'. 
A differentiated view could be considered for the creation of pools or lanes, depending on related processes and collaboration between the actors.

For the transformation of processes, we differentiate between tasks and sub-processes on the target side.
Processes with at least one aggregation or composition relationship to an underlying process becomes sub-processes (Rule 2).
In this example, we assume the assignment of multiple meta-types as shown in Table~\ref{table:metatypes}.
It will use the meta-link "Aggregate" to consider both relationships. 

A Business Process becomes a task if it does not aggregate any other sub-processes.
This task is embedded into the higher sub-processes of Rule 2, if any exist (Rule 3).
For this purpose, we use a relation with the type "Nested Element", which must be interpreted manually by the content interpreter when creating the BPMN model.

Finally, we want to transform the direct trigger relations between \textit{Business Processes} into \textit{Sequence Flows} (Rule 4). 
Since, we do not know in this context whether the processes have become a sub-process or a task, we reference both rules (2 and 3) and insert a placeholder for the target type.
Since the input conditions of both rules are mutually exclusive, only one returns the desired activity at a time.

\section{\uppercase{Conclusion and Future Work}}
\label{sec:conclusion_future_work}

Model transformation is a complex challenge due to the heterogeneity of languages and frameworks.
Especially when it comes to automated model generators, the difficulty lies in the many details of the enterprise models.
For this purpose, we have presented a concept that is language independent.
The system offers flexible description possibilities for model transformation using a rule-based approach with simple expressions.
It uses the idea of set theory and propositional logic for its description, so that only little previous knowledge is necessary for their use.
The established rules can be chained with each other.
In comparison to current approaches, the concept is easy to use and can be applied widely.
Based on the developed system, a natural, formal textual description language will be defined in the next step.

\bibliographystyle{apalike}
{\small
\bibliography{cleanbib}}

\end{document}